\documentclass[prl,showpacs,preprintnumbers,amsmath,amssymb]{revtex4}
\usepackage[dvips]{graphics}

\begin{document}

\title{Decoherence of massive superpositions induced by coupling to a quantized gravitational field}

\author{Vlatko Vedral}
\email{vlatko.vedral@physics.ox.ac.uk} 
\affiliation{Clarendon Laboratory, University of Oxford, Parks Road, Oxford OX1 3PU, United Kingdom and\\Centre for Quantum Technologies, National University of Singapore, 3 Science Drive 2, Singapore 117543 and\\
Department of Physics, National University of Singapore, 2 Science Drive 3, Singapore 117542}

\date{\today}

\begin{abstract}
We present a simple calculation leading to the quantum gravitationally-induced decoherence of a spatial superposition of a massive object in the linear coupling regime. The point of this calculation is to illustrate that the gravitationally-induced collapse could be of the same origin as any other collapse, i.e. due to the entanglement between the system (here a massive object) and its environment (in this case gravity, but it could well be the electromagnetic or any other field). We then point out that, in some cases, one has to be careful when concluding that matter-wave interference of large masses (to be quantified) would be prevented by their coupling to the gravitational field. We discuss how to experimentally discriminate between decoherence due to entanglement, decoherence due to classical dephasig as well as a genuine collapse of quantum superpositions (if such a process exists at all). 
\end{abstract}


\maketitle                           %display desi d

\section{Introduction} 

The superposition principle in quantum physics is, so far as all our experiments are concerned, universal. It applies to any degree of freedom and any pair of allowed quantum states (here we ignore superselection rules, whose origin may well be fundamental in some cases). How long a superposition can be maintained seems to be a question of how long the system can be insulated from various environmental influences, including an active application of various quantum error correcting protocols (i.e., whether superpositions are possible is not a fundamental question, but merely a question of money). 

However, when it comes to gravity being the actual environment of the system, things might at first sight appear to be different. First of all, we can never shield anything from gravity. Any object that has energy (and everything does) is acted upon by gravity. This includes gravity itself - ``gravity gravitates" - making the resulting classical theory of General Relativity non-linear. Secondly, we do not know how to treat gravity quantum mechanically so that it is difficult to apply the standard decoherence picture in which the environment becomes entangled with the system initially in a superposition thereby reducing the superposition to a mere classical mixture. Thirdly, there are various suggestions that gravity should not (or, may not have to?) be quantized, and that it may, in fact, be the ultimate cause of the collapse of quantum superpositions. There could well be more reasons why gravity is different, and none of them are necessarily logically or physically independent from each other. 

In this paper we would like to point out how a gravitational decoherence and collapse could simply be just like any other form of decoherence. Namely, their origin could be due to the entanglement created between the system and the quantized states of the gravitational field (assuming the linear regime). The mechanism is simple. We assume that a mass $m$ is superposed across a distance $\Delta x$ in some background gravitational field. The two different locations will lead to different interactions with the underlying gravitational field. The field will initially be assumed to be in a coherent state as is appropriate for the classical regime. If the mass is in one location, it displaces the gravitational coherent state one way, while if it is in the other location, it displace the field differently. The amount of displacement will be proportional to the ratio of the energy difference between two locations of the particle to the Planck energy, $\Delta E/E_P$, where $E_P=m_Pc^2$ and $m_P$ is the Planck mass. This way, the two massive spatial states become entangled to the gravitational field, which, when the field is traced out, leads to decoherence (just like any other type of decoherence in quantum physics). 

It is, however, also concievable, that the loss of coherence is due to the interaction with the classical gravitational field (though semi-classical models suffer from many inconsistencies which we will ignore in the present discussion). In this case, the field affects the phase between the two massive states, but does not become entangled to them (since it is now assumed to be classical and therefore incapable of exhibiting entanglement). This kind of noise is sometimes refered to as dephasing. We show how to discriminate this form of ``classical" dephasing from the genuine quantum decoherence. In summary, the answer will be via the effect called ``spin echo": classical dephasing can always, at least in principle, be reversed by acting only on the system; the decoherence due to the entanglement with the environment needs us to be able to act on the environment too, or to actively error-correct to keep the system disentangled from the environment. 

Neither entanglement-induced decoherence, nor classical dephasing are what is traditionally thought of as the objective collapse - in which gravity somehow forces a modification of quantum physics by leading to collapse, but not by either of the two (unitarity-compatible) methods described 
above. This objective collapse too can in principle be discriminated from other forms of decoherence. Interestingly, however, and as far as the dynamics of the system is concerned, the gravitational collapse can always be though of as a form of decoherence. This is perhaps not surprising given that any non-unitary, but completely positive, quantum evolution can be viewed as a unitary one from the perspective of the higher Hilbert space (assuming that collapse is a completely positive transformation). To discriminate a genuine collapse from this higher level decoherence processes would then require us to have the full knowledge and control of the environment in which case the relevant experiment is just an interference experiement that also involves the environment. While this may be hard in practice, it is certainly at least always possible in principle. 

Finally, we discuss the phenomenon of ``fake decoherence" where the gravitational entanglement to the massive superposition does not actually prevent us from observing interference. This is perhaps the most subtle message here, namely that not everything that looks like decoherence, actually leads to ``irreversible" decoherence. We will revisit matter wave diffraction experiments from this perspective. 

The literature on this topic is vast, however, frequently confusing and contradictory. In our view this mainly reflects the nature of the topic and the present lack of experiemental evidence regarding quantum effects in the gravitational field. The papers that are directly relevant for the linear gravity model we are using are \cite{Boughn,Weinberg,Dyson,Kay,Blencowe,CJR,MAVE,SOUG,Skagerstam}. It is worth bearing in mind that they are all differently motivated, however, they all have in common the linear model of gravity. The most relevant paper that addresses the classical gravitational dephasing is \cite{Brukner} and the literature on genuine collapse can be found here \cite{Karolyhazy,Diosi,Penrose}. The fact that semiclassical treatment of gravity is physically inconsistent was first clearly spelt out in \cite{Page} (see also \cite{Ford}), but we will ignore this fact and assume that gravity could in principle still be treated classically when coupling to quantum states of matter. The bias of the present author is that there is no such thing as classical noise, or classical anything, however, this prejudice does not affect any of the conclusions of the present work.

\section{Gravitational Decoherence Model}

We now present the main result, namely the entanglement induced decoherence of a massive superposition due to coupling to the quantized gravitational field in the linear regime. 
We pursue the analogy with the electromagnetic field and treat gravity in the linear regime in exctly the same way (see e.g. \cite{Boughn,Skagerstam}). By analogy, the gravity matter interaction Hamiltonian is given by:
\begin{equation}
H^G_{int} = \frac{1}{2} h_{\mu\nu} T^{\mu\nu} 
\end{equation}
where $T^{\mu\nu} \propto p_{\mu}p_{\nu}$
is the stress energy tensor of the single massive particle with momentum $p$ and $g_{\mu\nu} = \delta_{\mu\nu}+h_{\mu\nu}$. The quantized
gravitational field $h_{\mu\nu}$ is in the linear regime expanded as:
\begin{equation}
h_{\mu\nu} = \sum_\rho \int d^3k \{ a(k,s)\epsilon_{\mu\nu} (k,s) e^{ik_\lambda x^\lambda} + h.c.\}
\end{equation}
where $a(k,s)$ annihilates a graviton of momentum $k$ and spin $s$ and $\epsilon_{\mu\nu}$ is the graviton polarization. The notation $\phi = k_\lambda x^\lambda$ follows Einstein's summation convention and it represents the usual $kx-\omega t$ phase. We will work in the non-relativistic regime where only the $T_{00}$ component matters. This includes the component $mc^2$ as well as any other background potential (gravitational of otherwise) which contributes to the effective mass. This also implies that only the $h_{00}$ component of the field metric perturbation matters.

For our purposes we study a superposition of a massive object existing in two different locations which are therefore subjected to different local (gravitational or otherwise) fields.  Let us denote such a superposition, together with the state of the gravitational field, in the following manner:
\begin{equation}
(|\Psi_m (x)\rangle + |\Psi_m (x+\Delta x)\rangle)\otimes |\alpha\rangle 
\end{equation}
where $x$ is one location of the mass $m$ and $x+\Delta x$ the other one. Here $|\alpha\rangle$ is the initial state of the gravitational field, which would, in the static linear regime, simply be $\langle \alpha |\hat\Gamma |\alpha\rangle = \Gamma^i_{00} =-1/2h_{00,i}= \phi_G$. However, this is irrelevant as what matters are the dynamical degrees of freedom and the shift in the coherent state amplitude that they cause. This state subsequently evolves into
\begin{equation}
|\Psi_m (x)\rangle |\alpha + \delta \alpha \rangle + |\Psi_m (x+\Delta x)\rangle\otimes |\alpha-\delta \alpha \rangle \label{initial}
\end{equation}
where $\delta \alpha$ is the change of the coherent state due to coupling to gravity (to be given below). When the field is now traced out, the off-diagonal elements of the state of the particle are reduced by the amount $e^{-|\delta \alpha|^2}$. This is a time-dependent quantity and it represents an exponential amount of decoherence. Given that $\delta \alpha \propto \Delta E/E_P$ we find that the rate of decoherence to scale as $\gamma = |\delta \alpha|^2/t\propto (\Delta E/E_P)^2$. 

In order to obtain an estimate of the decoherence rate, we use the full linear regime Hamiltonian where the gravity-matter interaction Hamiltonian is:
\begin{equation}
H^{G}_{int} =\int_x dx \int_{k} dk \psi^{\dagger} (x)\psi (x) g_k (x) (a_{k\sigma}e^{i\omega_k t}+a^\dagger_{k\sigma}e^{-i\omega_k t})
\end{equation}
where, 
\begin{equation}
g_k = \frac{V(x) x^2\sqrt{16\pi G\hbar}}{4c^3\sqrt{V}}  \omega_k^{3/2} \; .
\end{equation}
We have ignored the $mc^2$ part of $T_{00}$ as it is the same for both states of the mass and will therefore not affect the relative phase between them. For simplicity, we have also surpressed the polarization degree of freedom and we have assumed that $x_ix_j=x^2$ since we are only interested in a rough estimate. Note that while the EM coupling is in the lowest order dipole, the gravitational coupling happens to be quadrapole and hence the scaling with $x^2$. Since the massive particle is assumed to be near stationary, the free part of the Hamiltonian is given by
\begin{equation}
H_{0} = \int dk (\hbar c k) a^\dagger_{k\sigma} a_{k\sigma}
\end{equation}
which is just the energy of the gravitational field. We will work in the rotating frame defined by $H_0$. Here we have that the annihilation and creation operators precess at the rate:
\begin{eqnarray} 
a_n & \rightarrow & a_n e^{i\omega_n t} \\
a_n^{\dagger} & \rightarrow & a_n^{\dagger} e^{-i\omega_n t} \; ,
\end{eqnarray}
and therefore the evolution due to the coupling becomes:
\begin{eqnarray}
U  =   \exp  \left\{  -i\int dx \psi^{\dagger}(x)\psi (x)\int_k dk g_k  \int_0^T(a_m e^{i\omega_m t} + a^{\dagger}_m e^{-i\omega_m t})dt \right\}
\end{eqnarray}
The evolution acts like a control gate, in the sense that it kicks the gravitational field only at the locations where the mass is present. Also, each momentum evolves independently of other momenta, and therefore, the relevant computation reduces to 
\begin{equation}
\exp \left\{-i g_k \int_0^T(a_k e^{i\omega_k t} + a^{\dagger}_k e^{-i\omega_k t})dt \right\}| \alpha_k \rangle \; ,
\end{equation}
where $T$ is the total time of evolution. This then allows us to calculate the shift of the gravitational coherent state and we readily obtain
\begin{equation}
| \alpha_k +\delta \alpha_k\rangle = |\alpha_k -i\frac{g_k}{\omega_k}(e^{i\omega_k T} -1)\rangle
\end{equation}
for each momentum state. The decoherence rate, when summed up over all momenta, then becomes: 
\begin{equation}
\frac{\gamma}{T} = \int_k dk g^2 (k) \rho (k) \frac{\sin^2 (\omega_k T) }{\omega^2_k} \; .
\end{equation}
The rate of decoherence depends on a number of assumptions, such as the relationship between the time of evolution and the relevant frequencies as well as the assumptions regarding the density of states $\rho (k)$ and the upper frequency cut-off $\Omega$ (see \cite{Ekert} for a detailed discussion). We can even include the effects of a finite temperature of the gravitational environement (all this requires in the above calulation is to assume that the field is in a thermal mixture of coherent states, as well as of course that we are in the usual Borm-Markov regime). The final result is straightforward to anticipate using a simple dimensional analysis (which more detailed calculations corroborate as in \cite{Ekert}):
\begin{equation}
\gamma \approx \left(\frac{\Delta E}{E_P}\right)^2 \left( \frac{\Delta x}{c} \right)^n \frac{kT}{\hbar} \Omega^n
\end{equation} 
where $n$ is some positive power that depends on the above mentioned details, $\Delta E$ is the energetic difference between the two massive particle configurations, $E_P=m_Pc^2$ and $T$ is the temperature. Note that, if the upper frequency cutoff is given by $\Omega = \delta x/c$, we recover the result in \cite{Blencowe}.

\section{Discussion: Genuine Quantum Gravity Versus Classical Decoherence and Collapse}

What kind of rates are expected in practice given this formula? An electron superposed across two energy levels in an atom should be stable over a long period of time. On the other hand, a much more massive object, superposed across larger distances might decohere more rapidly.
Interestingly, if we assume that $\Delta E=mc^2$, the above rate at zero temperature
$\gamma = \left(\frac{\Delta E}{E_P}\right)^2 \left( \frac{\Delta x}{c} \right)^n \Omega^{n+1}$ reduces to the Penrose collapse formula. This is consistent with Penrose's interpretation that the fluctuations of the gravitational field produce an energy which, when divided by $\hbar$ gives the rate of the collapse. Here we see that there is a way of formally arguing for the gravitationally-induced collapse, but which is in no way different to any other form of quantum decoherence. Namely, imagine that the gravitational field is described by the quantised Christoffel symbols, $\hat \Gamma = \partial \hat g$. Then, the fluctuations in the gravitational field are \cite{Anandan}: 
\begin{equation}
\Delta^2 \Gamma = \int d^3 x \langle \alpha| \hat \Gamma^2 |\alpha\rangle - \langle \alpha| \hat \Gamma |\alpha\rangle^2
\end{equation}
Here gravitational decoherence is then seen as being induced by the fluctuations of the quantized gravitational field. This is the same as in the electromagnetic case of say the vacuum inducing spontaneous emission, or a more general dephasing and other such phenomena (see \cite{Ford} for the differences between the semi-classical and full quantum treatments). 

It is worth mentioning, however, that the above process does not involve dissipation. The energy of the massive superposition does not change, namely the diagonal elements are not affected by the decoherence. The linear quantum gravity model can also be used to obtain dissipation (say through spontaneous emission of gravitons). The standard calculation for the quadrapole emission gives us the rate:
\begin{equation}
\frac{dE}{dt} = -\frac{Gm^2a^4\omega^6}{c^5} =- \gamma_s \hbar \omega
\end{equation}
and this result can be also be obtained for say an electron making a transition within an atom by emitting a graviton in the linear regime \cite{Boughn,Weinberg,Dyson} (the formula is the same as the dipole emision of light, providing we make the substitutions $e\rightarrow \sqrt{G}m$ and $a\rightarrow a^2)$. The rate of emission $\gamma_s$ would be so small that the half life would be orders of magnitude bigger than the age of the Universe. As was pointed out by a number of people including the present author \cite{MAVE,SOUG}, the fact that gravitons may not be observable does not imply that we cannot confirm the quantum nature of gravity.

Now we reach the crucial part of the discussion. Can we discriminate a genuine collapse from various possible entanglements with the quantum gravitational field? Yes, but only if we have access to other measurements of the field. If all we can do is measure the rate of decoherence of the superposed mass, then the origin of the decoherence cannot be established. What can be done is to discriminate an objective collapse from classical dephasing. Namely, imagine that the phases of the two states in the superposition evolve at different rates (as would be the case with when an atom is superposed at two different heights in Earth's gravitational field as in \cite{Brukner}). Then at some time $t$ the state would be 
\begin{equation}
e^{i\phi (x)t}|\Psi_m (x)\rangle + e^{i\phi (x+\Delta x)t} |\Psi_m (x+\Delta x)\rangle 
\end{equation}
Now, if there relative phase $\phi (x+\Delta x)-\phi (x)$ is large (compared to the time of measurement, say) then this will lead to the collapse of the interference between the two components, which is what we called classical decoherence. This decoherence, however, is only apparent and can actually be ``reversed". The standard technique is the spin echo, which just swaps the two elements of the superposition half way through the experiment. This equalises the two phases at the end, which therefore just become a global phase. Therefore the classical dephasing, unlike the objective collapse, can be undone. The key facts are that a genuine collapse can never be reversed (by definition), the classical dephasing can be reversed by acting on the system only and the decoherence due to entanglement can be reversed if we have access to and control of the environment causing the decoherence through coupling to the system.

Finally, we discuss the notion of fake decoherence. When a neutron undergoes interference
as in \cite{COW}, its spin couples to the neighbouring spins. Since in the interferometer the neutron exists in a superposition of 
two different spatial locations, the neighbouring spins are in two different states which are entangled with the spatial states of the neutron. If we were to trace out the environmental spins, the neutron spatial supperposition would seemingly decohere. This however does not prevent us from observing neutron interference! It is so because when the two arms of the interferometer are recombined to measure the interference, the two environmental states also merge into effectively the same state. This is why despite the fact that the neutron was fully entangled with envorinmental spins inside the interferometer we are still able to observe interference at the end. 

The same would occur with a massive superpostion and the gravitational field which would only be responsible for decoherence if the two states of the field did not return back to the same state when the two massive states were recombined to interfere. Therefore, a massive superposition would not collapse due to this, and what is required, for instance, is an emission of a graviton which would for all practical purposes be irreversible (just like a spontaneous emission of a photon).  

This naturally leads us to revist the scenario in \cite{Baym} where a massive particle undergoing a double slit interference interacts with another massive particle such that the resulting configuration reveals which slit the particle has gone through (thereby destroying interference). The conclusion of that paper is that in order for classical gravity not to spoil interference, the interference fringes would have to be smaller than the Planck length. This result too, can be modeled with the fully quantized field. Briefly, the logic is as follows. The energy difference between the two configurations where the mass is at the two slits should be at least as big as a single graviton energy (whose wavelength should not be smaller than the distance between the slits in order not to destroy the interference). This graviton would be emitted by the process of gravitational bremsstrahlung \cite{CJR}. Therefore, $Gm^2 d/r^2 \geq hc/d$, where $r$ is the distance between the two masses and $d (<<r)$ is the slit separation. This is exactly the condition in the paper \cite{Baym} that leads the authors to conclude that the size of the interference fringes would be below the Planck length (and therefore presumably unobservable). 

The fact that a quantum gravitational argument could be used to reach the same conclusion reinforces the idea that any classical collapse (as in the paper \cite{Baym} or indeed in the paper \cite{Brukner}) can be reproduced quantum mechanically. We therefore have to be careful regarding our conclusions about the quantum nature of gravity just based on the bare fact that a certain experiment is (or isn't) capable of demonstrating any quantum interference effects. After all, in a fully quantum universe, the classical world exist only because of the entanglements between various subsystems whose quantum interference effects are therefore impeded due to the overall higher level quantumness  \cite{Joos}. Even the classical world is classical only because the universe is (at least to a high enough degree) quantum.

\textit{Acknowledgments}: The author acknowledges many stimulating discussions with Chiara Marletto. This resaerch has been supported by the Centre for Quantum Technologies in Singapore.

\end{document}